\documentclass[aoas,preprint]{imsart}

\usepackage[normalem]{ulem}
\newcommand{\strike}[1]{}

\usepackage{xcolor}
\newcommand{\blue}[1]{#1}
\usepackage{float}

\RequirePackage{amsthm,amsmath,amsfonts,amssymb}
\RequirePackage[authoryear]{natbib}
\RequirePackage{graphicx}

\startlocaldefs

\newtheorem{define}{Definition}

\newcommand{\ps}{\{5,8,10\}}
\newcommand{\n}{381 }
\newcommand{\pp}{295 }
\newcommand{\ppp}{2663 }
\newcommand{\lo}{0.0210 }
\newcommand{\hi}{0.9291 }

\newcommand{\singp}{63.88\% }

\newcommand{\cmmnt}[1]{}
\newcommand{\Es}[2]{\mathbb{E}_{#1}\left[#2\right]}

\newcommand{\I}{\mathbf{I}}
\newcommand{\J}{\mathbf{J}}
\renewcommand{\P}{\mathbb{P}}
\renewcommand{\Pr}[1]{\mathbb{P}\left(#1\right)}

\newcommand{\R}{\mathbb{R}}

\renewcommand{\v}{\mathbf{v}}
\newcommand{\twop}{2^p}
\newcommand{\D}{\mathbf{D}}
\newcommand{\Ical}{\mathcal{I}}
\newcommand{\K}{\mathbf{K}}
\newcommand{\Rbf}{\mathbf{R}}
\newcommand{\Rcal}{\mathcal{R}}
\newcommand{\Rast}{\Rcal_\ast}
\newcommand{\T}{\mathbf{T}}
\newcommand{\U}{\mathbf{U}}
\newcommand{\W}{\mathbf{W}}
\newcommand{\X}{\mathbf{X}}

\newcommand{\one}{\mathbf{1}}
\newcommand{\zero}{\mathbf{0}}

\newcommand{\m}{\mathbf{m}}
\newcommand{\ms}{\tilde{\m}}

\newcommand{\w}{\mathbf{w}}
\newcommand{\x}{\mathbf{x}}
\newcommand{\y}{\mathbf{y}}
\newcommand{\z}{\mathbf{z}}
\newcommand{\zs}{\tilde{\z}}
\newcommand{\zse}{\tilde{z}}
\newcommand{\zop}{\{0,1\}^{p}}
\newcommand{\zotp}{[0,1]^{2^p}}

\newcommand{\diag}[1]{\mbox{diag}\left(#1\right)}
\newcommand{\expl}[1]{\mbox{expl}\left\{#1\right\}}

\newcommand{\KL}[2]{\mbox{KL}\left(#1||#2\right)}

\newcommand{\vbeta}{\boldsymbol\beta}
\newcommand{\vgamma}{\boldsymbol\gamma}

\newcommand{\hs}{\hspace{1em}}

\renewcommand{\t}{5000 }

\newcommand{\twoprop}{0.739 } 

\newcommand{\threeprop}{0.162 } 
\newcommand{\compthree}{0.904 } 

\newcommand{\IQR}{(0.636,0.871) }
\newcommand{\eqtwo}{6.355\% }
\newcommand{\IQRthree}{(0.868,0.991) }
\newcommand{\IQRapx}{(0.607,0.754) }

\newcommand{\medfive}{0.683}
\newcommand{\medeight}{0.679}
\newcommand{\medten}{0.697}

\newcommand{\rastmed}{0.774 }
\newcommand{\rastiqr}{(0.713,0.819)}
\newcommand{\rastlo}{71.3\% }

\endlocaldefs

\begin{document}

\begin{frontmatter}
\title{Information content of high-order associations of the human gut microbiota network}
\runtitle{Information content of high-order associations}

\begin{aug}
\author[A]{\fnms{Weston D.} \snm{Viles}\ead[label=e1,mark]{weston.viles@maine.edu}},
\author[B]{\fnms{Juliette C.} \snm{Madan}\ead[label=e2,mark]{Juliette.C.Madan@hitchcock.org}},
\author[C]{\fnms{Hongzhe} \snm{Li}\ead[label=e3]{hongzhe@upenn.edu}},
\author[B]{\fnms{Margaret R.} \snm{Karagas}\ead[label=e5,mark]{margaret.r.karagas@dartmouth.edu}}\and
\author[A]{\fnms{Anne G.} \snm{Hoen}\ead[label=e6,mark]{anne.g.hoen@dartmouth.edu}}
\address[A]{Department of Biomedical Data Science, Geisel School of Medicine at Dartmouth \printead{e1,e6}}

\address[B]{Department of Epidemiology, Geisel School of Medicine at Dartmouth \printead{e2,e5}}

\address[C]{Department of Biostatistics and Epidemiology, Perelman School of Medicine, University of Pennsylvania \printead{e3}}

\end{aug}

\begin{abstract}
The human gastrointestinal tract is an environment that hosts an ecosystem of microorganisms essential to human health.  Vital biological processes emerge from fundamental inter- and intra-species molecular interactions that influence the assembly and composition of the gut microbiota ecology.
Here we quantify the complexity of the ecological relationships within the human infant gut microbiota ecosystem as a function of the information contained in the nonlinear associations of a sequence of increasingly-specified maximum entropy representations of the system.  Our paradigm frames the ecological state, in terms of the presence or absence of individual microbial ecological units that are identified by amplicon sequence variants (ASV) in the gut microenvironment, as a function of both the ecological states of its neighboring units and, in a departure from standard graphical model representations, the associations among the units within its neighborhood.
We characterize the order of the system based on the relative quantity of statistical information encoded by high-order statistical associations of the infant gut microbiota.
\end{abstract}



\end{frontmatter}


\section{Background}

The ecological relationships of the microbial ecosystem of the human gut are influenced by fundamental molecular interactions among microorganisms and underpin the formation of complex, robust communities that correspond to important biological functions for its host \citep{Backhed2005,Baumler2016,Trosvik2010,Kahrstrom2016}. The fundamental relationships among mutually-coupled microorganisms of the gut ecosystem possibly involve multiple  distinct ecological units, i.e., microbes existing and participating within the ecosystem that are identified as amplicon sequence variants (ASV) in commensal, symbiotic, and pathogenic interactions \citep{Haque2017,Shoaie2013}. Statistical analysis of the states of this system may reveal the ecological associations, i.e., combinations of ecological relationships among the incident ecological units that manifest from these fundamental relationships.
An ecological relationship is basically quantified by its order,  i.e., the number of  distinct ecological units involved in the association, and the direction and magnitude of its corresponding statistical association measurement, e.g., correlation.  Graphical models and network analysis methods are often-used tools for representing the observed second-order, i.e., pairwise, ecological relationships of the gut ecosystem \citep{Freilich2010,Layeghifard2017,Ramette2007} and, yet, are not sufficiently descriptive to express the stability-diversity paradox observed in nature  \citep{Ives2007,Bairey2016}.
Associations have commonly been estimated with pairwise correlation \citep{Mandakovic2018,Aas2005} despite its technical limitations   \citep{Aitchison1981,Poudel2016} or by partial correlation through generalized linear models  \citep{Faust2012a}, including logistic regression \citep{vandenBergh2012}.  Network motifs and clusters are frequently identified subsequently as proxy higher-order associations  \citep{Milo2002,Alon2007,Faust2012b} although, as we demonstrate in the following, this paradigm of communities as composites of pairwise relationships cannot fully capture the breadth of complex ecological associations in the system.

Early proponents \citep{Abrams1983,Billick1994} for the existence of high-order interactions, i.e., involving more than two ecological units, in ecological systems described the complex relationships in terms of one entity modifying the nature of an interaction in the system.  Detecting the nonlinear relationships that comprise a high-order association is a long-standing statistical problem \citep{Case1981}.  More recently, the functionality of high-order interactions has been inferred to include the promoting of stability and diversity in ecological communities   \citep{Grilli2017,Levine2017,Friedman2017}. Methodology for the integration of more than two operational taxonomic units (OTUs) \citep{Mandakovic2018} in a regulatory  triplet model \citep{Tsai2015} and estimation of dynamic networks that evolve according to a state space model \citep{Chen2017}, a special case of dynamic Bayesian networks \citep{Durbin2001}, have been used in constructing microbial network models.  Within the microbial communities of the human gut, higher-order interactions may occur on account of competing enzyme-antibiotic production  \citep{Kelsic2015} or environmentally adaptive trophic interactions  \citep{Beckerman1997} in which ancillary species influence pairwise interactions to support and regulate the diversity of multi-species communities  \citep{Poisot2015,Wootton1994}.  Such high-order associations have been shown to influence host fitness traits in the \textit{Drosophila melanogaster} fruit fly, the microbiome of which consists of few combinations of microorganisms \citep{Gould2018} that are feasibly enumerated.

A common mathematical reduction of the state of an ecosystem is the binary vector of zero-one occurrence indicators reflecting the absence-presence of the distinct ecological units of an ecosystem \citep{Ramette2007,Mandakovic2018,Callahan2017}.  The statistical properties of this binary state vector are derived from the probability distribution characterizing the likelihood of ecological states.  On ecological subsystems, we identify a class of low-rank approximations of this empirical probability distribution, commonly known as the maximum entropy distributions \citep{Jaynes1957}. With each subsystem consisting of a fixed number of distinct ecological units, we quantify the relative information gains associated with statistical representations of increasingly higher rank.  As a corollary, we subsequently lower bound the maximal order of association in the subsystems.  Based on a sample of $\n$ infants, with a single observation corresponding to each, we demonstrate the broad existence of high-order
ecological associations in sub-ecosystems of the human gut microbiota.  Our conclusion supports modern ecological theory on the combinatorial nature of mutual influence among biological entities \citep{Bairey2016}.  Consideration of higher-order statistical associations is pertinent to the accurate prediction/detection of abnormal states of the human microbiota.

The methodology we developed is a general procedure for estimating a low-rank approximation of the statistical distribution of states in a binary system and for characterizing, in terms of order, the complexity of its associations.  We understand ``complexity'' in the context of a complex system and, in particular, the statistical information attributed to nonlinear associations that characterize such systems.  We represented the observed states of the binary system (composed of zero-one ecological occurrence indicator variables) with a sequence of progressively-specified maximum entropy models indexed by order.  A maximum entropy model $\mathbb{P}(\x)$ takes the form
$\mathbb{P}(\x) \propto \exp\{\sum_j \lambda_jf_j(\x)\}$ \citep{Zdravko2011}, where the $\lambda_j$ are constants and $f_j(\x)$ are conserved quantities, i.e., statistics / functions of the data.  This probability distribution is increasingly specified with the inclusion of subsequent conserved quantities that are the statistics which further constrain the model.  The increasingly-descriptive statistical models in this sequence potentially facilitate, in turn, an increase in the predictability of states as quantified through entropy-based measures \citep{Kullback1951}. Specifically, we measured the relative statistical information attributed to the high-order statistical associations relative to those associations of lower-order in the sequence of estimated low-rank approximations of the statistical distribution of states.  Our methodology implicitly quantifies the statistical information gained through representations of the high-order associations of the gut microbiota ecosystem and thereby demonstrates that third-order ecological relationships are abundant and important for characterizing the statistical properties of the subsystems. 

\section{Methods}

We describe the sample collection and the numerical methods of our procedure in the following subsections.

\subsection{Sample collection, sequencing, and processing}

Mothers aged 18 to 45 years participating in the New Hampshire Birth Cohort Study at Dartmouth provided infant stool samples that were collected at regularly scheduled maternal six-week postpartum, follow-up visits (mean: 46 days, median: 44 days, range: 14-153 days, s.d.: 13.7 days). Institutional review board approval was obtained at Dartmouth with yearly renewal.  Subjects provided written informed consent to participate on behalf of themselves and their infants.  A total of $374$ mothers participated in the present study, seven of whom participated with two singleton sibling infants on separate occasions.  A total of $n=381$ unique infants provided one stool sample each to the present study.

Stool was aliquoted in sterile tubes and frozen at ${-80}^\circ$C within 24 hours of receipt. Samples were thawed and DNA was extracted using the Zymo DNA extraction kit (Zymo Research). The quantity and purity of the DNA were determined by OD260/280 nanodrop measurement. Illumina tag sequencing of the 16S rRNA gene v4-v5 hypervariable region and initial quality control was performed at the Marine Biological Laboratory in Woods Hole, Massachusetts.
Quality control procedures eliminated sequences containing more than one ambiguous nucleotide, removed sequences with a length outside of the expected distribution, and eliminated chimeric reads using the UCHIME algorithm  \citep{Edgar2011} de novo and with reference within the USEARCH program \citep{Edgar2010}.

Sequences were processed using the DADA2 sequence processing pipeline (v.1.6.0) \citep{dada} to infer the amplicon sequence variants (ASVs) present and their relative abundances across samples. Sequencing and sequence read processing were done using established methods that have been previously described \citep{Coker2019,Newton2015,Huse2014}.
\strike{We identified} \blue{The sequencing process identified} $\ppp$ ASVs from $\n$ infant stool samples.\footnote{DNA sequence data are available at the Genbank Sequence Read Archive (https://www.ncbi.nlm.nih.gov/genbank/) under accession number PRJNA296814.}  \blue{We examined our protocol extensively and found no evidence of batch effects} \citep{Antosca20}.  \blue{On the $\log_{10}$ scale, the read counts had an average of $4.951$ and standard deviation of $0.282$.  Again, on the $\log_{10}$ scale, the read counts had a median of $4.995$ with the $0.025$ and $0.975$ quantiles as $Q_{0.025}=4.331$ and $Q_{0.975}=5.276$.}

The ASV abundances recorded in the $\n\times \ppp$ ASV table were transformed to the $\n\times\ppp$ binary matrix with the $(i,j)$ element equal to one if the abundance of the $j^{th}$ ASV in the $i^{th}$ sample exceeded the detection limit, i.e., the corresponding entry in the ASV table was positive, and otherwise zero, for $i\in\{1,2,\ldots,\n\}$ and $j\in\{1,2,\ldots,\ppp\}$.  We computed the univariate information entropy $H(\hat{p}_j)=-\hat{p}_j\log(\hat{p}_j)-(1-\hat{p}_j)\log(1-\hat{p}_j)$
of occurrences for the $j^{th}$ ecological unit, where $\hat{p}_j$ was the occurrence rate (sample proportion) of the $j^{th}$ ASV binary ecological occurrence variable, for $j\in\{1,2,\ldots,\ppp\}$. Subsequently, we eliminated those with low entropy via the elbow method \citep{Hastie2001}.
This amounted to our retaining $\pp$ ASVs in the $\n\times\pp$ binary matrix $\D$ with occurrence rates in the interval $[\lo,\hi]$ and focused our exploration for high-order associations in high-entropy components of the ecosystem.\footnote{The greatest ASV occurrence rate in the data set was $0.9291$ and did not exceed the upper threshold of $1-0.0210 = 0.9790$ of our entropy filter.  Accordingly, no high-rate ASVs were discarded from our analysis.}	The discarded low-rate \strike{and high-rate} ASVs are predominated ($\singp$) by ASVs with a single occurrence in the data set and are easily predictable in their own rights.  \blue{This may be asserted since a binary ecological variable that has a relative frequency of $\epsilon$, for some small $\epsilon$, is deterministically  predicted as either one or zero, respectively, with error rate $\epsilon$.  Accordingly, in this investigation, we focus our attention on those ecological variables of lesser first-order predictability since these variables will control the lack of predictability of the entire system.}  \blue{Subsequent to the application of the entropy filter, the read counts of the $n=\n$ samples on the $\log_{10}$ scale have an average of $4.918$ and standard deviation of $0.287$.  Again, on the $\log_{10}$ scale, the read counts have a median of $4.969$ with the $0.025$ and $0.975$ quantiles as $Q_{0.025}=4.295$ and $Q_{0.975}=5.228$.}  \blue{We note that the minimum entropy criterion for inclusion in the analysis required that an ASV have a non-zero in at least eight samples.  That is, the maximum number of non-zero reads an ASV may have recorded and, nevertheless, be excluded from further analysis is seven.  It follows that all excluded ASVs were undetected in at least $381-8=373$ $(97.9\%)$ samples.}

\blue{The binary transformation that we imposed on the data, maps a null abundance to zero.  Conversely, a positive abundance is mapped to one.  This transformation is historically fundamental in ecological community description \citep{Mackenzie04} and is commonly used to approximate the interactions that drive community assembly \citep{BarMassada2015,MoruetaHolme16} and modify community organization \citep{Kay18}.  Our descriptive analysis of the high-order nature of ecological associations of the infant gut microbiome has identified targets for specific inference of tropic and other interspecific interactions of organisms within the the gut microenvironment.} 

\blue{An analysis of the correspondence between the high-order associations discovered in the binary system of occurrence variables of the gut microbiota ecosystem in the manner we have presented here and those derived experimentally, to our knowledge, has not been conducted on a similarly large scale to the present analysis.  In our forthcoming work, we intend to differentiate the ecological information attributed to high-order statistical associations among the binary system, as we have derived here, and the corresponding set of high-order statistical associations among the original set of ASV relative abundances.  Inference of the nature of the population interactions that our descriptive analysis has enumerated and their biological and ecological functions remains in the work ahead of us and for those involved in microbiota ecology in which content-knowledge of the species involved may be incorporated for further detail.}

\subsection{Low-rank Approximation}

The occurrence of $\pp$ distinct ecological units in an observation of the gut ecosystem is a binary (one if present, zero if absent) state vector $\x\in\{0,1\}^{\pp}$ in a sample space of $2^{\pp}> 10^{88}$ states.  We concentrate on sub-ecosystems consisting of a fixed number of distinct ecological units $p$, for $p\in\ps$, for which the quantities of statistical information encoded by associations can be feasibly computed and recorded over a multitude of instances.
Statistical properties of the $p$-length binary random vector $\x\in\zop$ are functions of its probability distribution $\P:\zop\mapsto[0,1]$.  Let $\x_k\in\{0,1\}^p$ be the binary representation of the number $k$, for $k\in\{0,1,2,\ldots,\twop-1\}$, and define the $\twop\times p$ matrix $\X$ to have $k^{th}$ row $\X_{k\cdot}=\x_k$.  The sample space of all $p$-length binary vectors $\x\in\zop$ is the union of the row vectors of $\X$.

Define $\z\in\R^{\twop}$ with $\|\z\|_1=1$ as the probability vector representing the likelihood of states $\x\in\zop$ to have components
\begin{align*}
z_k &= \Pr{\x=\x_k},
\end{align*}
for $k=0,1,\ldots,\twop-1$.  Let $u(\x) = \{i:x_i=1\}$ be the indexes of components in $\x$ equal to 1 and define the sequence of indicator functions
\begin{align*}
T_k(\x) = 1\{u(\x)\subseteq u(\x_k)\}.
\end{align*}
The moments $m_k$, for $k\in\{1,2,\ldots,\twop-1\}$, of $\z$ are enumerated as
\begin{align*}
m_k=\Es{\z}{T_k(\x)} &= \sum_{j=0}^{\twop-1}z_jT_k(\x_j).
\end{align*}
By the fact that $\z$ is a probability vector, it is straight-forward that $m_0=1$.  We construct the $\twop\times(\twop-1)$ zero-indexed matrix $\T$ with elements
\begin{align*}
\T_{jk} &= T_k(\x_j)
\end{align*}
and note that the moments $\m=(m_1,\ldots,m_{\twop-1})^\prime\in[0,1]^{\twop-1}$ of $\z$ satisfy
\begin{align*}
\m &= \T^\prime\z.
\end{align*}
This illustrates the bijection between the state probability vector $\z$ and the moment sequence $\m$.

Provided that $\z$ is strictly positive, the statistical distribution of states $\x\in\zop$ may be represented as the Gibbs distribution
\begin{align}\label{gibbs}
z_k &= \Pr{\x=\x_k|\vgamma} = \exp\left\{\sum_{j=1}^{\twop-1}\gamma_jT_j(\x_k)-\log(Z(\vgamma))\right\},
\end{align}
for some $\vgamma\in\R^{\twop-1}$, where $Z(\vgamma)$ is the partition function
\begin{align*}
Z(\vgamma) &= \sum_{k=0}^{\twop-1}\exp\left\{\sum_{j=1}^{\twop-1}\gamma_jT_j(\x_k)\right\},
\end{align*}
i.e., the normalizing constant of the probability distribution.
With $\expl{\cdot}$ as the element-wise exponential function, we have
\begin{align}\label{zrep}
\z &= \expl{\T\vgamma-\log(Z(\vgamma))\one},
\end{align}
where $\mathbf{1}$ is the length $2^p$ vector of ones. Define $g:\R^{\twop-1}\mapsto\R^{\twop}$as $g(\vgamma)=\expl{\T\vgamma-\log(Z(\vgamma))\one}$ and note that the columns $\T$ are the basis vectors of the nonlinear transformation $g(\cdot)$ from $\R^{\twop-1}$ to $\zotp$.


A low-rank approximation of the probability vector $\z$ will exploit any redundancy in the moment sequence $\m$.
We classify moments of $\z$ according to the number $|u(\x)|$ of active states, i.e., number of ones in $\x$.  To that end,
define the increasing sequence of index sets
\begin{align*}
\I_d &= \{k:|u(\x_k)|\leq d\},
\end{align*}
for $d\in\{1,2,\ldots,p\}$, and note that $|\I_d|=t_d$, where $t_d=\sum_{i=1}^d\binom{p}{i}$. Correspondingly, let $\T_d$ be the $\twop\times t_d$ matrix constituted by the columns of $\T$ corresponding to the indexes in $\I_d$, for $:d\in\{1,2,\ldots,p\}$.  Finally, we define the increasing subsets $\U_d\subseteq[0,1]^{\twop}$ according to
\begin{align*}
\U_d = \{\y\in\R^{\twop}:\exists\vbeta\in\R^{t_d}\mbox{ s.t. }\y=g_d(\vbeta)\},
\end{align*}
where $g_d(\vbeta)=\expl{\T_d\vbeta-\log(Z(\vbeta))\one}$, for $d\in\{1,2,\ldots,p\}$.  That is, $\U_d$ is the image of all $\vbeta\in\R^{t_d}$ under the nonlinear transformation $g_d(\vbeta)$.
\bigskip

\begin{define}
	The binary system with states $\x\in\zop$ and state probability vector $\z\in\zotp$ is a $d^{th}$-order system if 
	\begin{align*}
	\z\in \U_d\mbox{ and }\z\notin \U_{d-1},
	\end{align*}
	for some $d\in\{2,3,\ldots,p\}$.
\end{define}

\subsection{Model Identification}

Let $\vbeta\in\R^{t_d}$ and define $\z_d=g_d(\vbeta)$, for some $d\in\{2,3,\ldots,p\}$.  The cross-entropy $H(\z,\z_d)$ from $\z_d$ to $\z$ is
\begin{align*}
H(\z,\z_d) &= -\sum_{k=0}^{\twop-1}z_k\log(z_{dk})\\
&= -\z^\prime\left[\T_d\vbeta-\log(Z(\vbeta))\one\right]\\
&= -\m_d\cdot\vbeta+\log(Z(\vbeta)),
\end{align*}
where $\m_d=\T_d^\prime\z$.	If $\z_d=\z$, then there is no statistical information lost in representing $\z$ with $\z_d=g_d(\vbeta)$.
In this case, $H(\z,\z_d)=H(\z)$ is the entropy of $\z$ and the predictability of states encoded in $\z$ has been captured in the $d^{th}$-order statistical representation $\z_d=g_d(\vbeta)$.  The gradient of $H(\z,\z_d)$ with respect to the parameter vector $\vbeta$ is
\begin{align*}
\frac{d}{d\vbeta}H(\z,\z_d) &= \frac{d}{d\vbeta}H(\z,g_d(\vbeta)) =  -\T_d^\prime\z+\frac{d}{d\vbeta}\log(Z(\vbeta)).
\end{align*}
Since $\frac{d}{d\beta_j}Z(\vbeta)=\Es{\z_d}{T_j(\x)}$, the gradient takes the form
\begin{align*}
\frac{d}{d\vbeta}H(\z,\z_d) &= \T_d^\prime(g_d(\vbeta)-\z).
\end{align*}
The system of equations
\begin{align}\label{score}
\zero &= \T_d^\prime(g_d(\vbeta)-\z)
\end{align}
are precisely those which identify the $d^{th}$-order maximum entropy distribution approximating the probability vector $\z$.  That is, all moments of at most $d^{th}$-order are conserved in the solution probability vector $\hat{\z}_d$.  The remaining moments are left unconstrained.

The $d^{th}$-order low-rank approximation $\hat{\z}_d=g_d(\hat{\vbeta})$ of $\z$, for solution vector $\hat{\vbeta}\in\R^{t_d}$ and based on the column vectors of $\T_d$ and subject to the constraint $\|\hat{\z}_d\|_1=1$, satisfies the system in Equation \eqref{score}.
Since the Hessian matrix of $H(\z,g_d(\vbeta))$ is positive definite for $\vbeta\in\R^{t_d}$, the solution $\hat{\vbeta}$ to the system in Equation \eqref{score} obtains the minimal cross-entropy to $\z$ from any probability distribution $\z_d\in\U_d$ and may be computed with a gradient descent procedure (see Appendices \ref{app:grad} and \ref{app:cv}).

If $\z\in\U_d$, then $\z_d=\z$ and the order of the system with state probability distribution vector $\z$ is at most $d^{th}$-order.  More precisely, the Kullback-Leibler divergence
\begin{align*}
\KL{\z}{\hat{\z}_d} &= -\sum_{k=0}^{\twop-1}z_k[\log(\hat{z}_{dk})-\log(z_k)],
\end{align*}
the relative entropy from the $d^{th}$-order probability distribution $\hat{\z}_d$ approximating the probability vector $\z$, equals zero only when $\hat{\z}_d=\z$.  In reference to Definition 1, if $\KL{\z}{\hat{\z}_d}=0$, for some $d\in\{1,2,\ldots,p\}$, then $\z\in\U_d$ and, more generally, $\KL{\z}{\hat{\z}_{d^\prime}}=0$, for $d^\prime\in\{d,d+1,\ldots,p\}$.

\subsection{Approximating the Probability Distribution of the Data}

In practice, samples $\x^{(1)},\x^{(2)},\ldots,\x^{(n)}$ are observed instances of the infant gut ecosystem and $\z$ is directly estimated with the empirical probability distribution $\zs$ of relative frequencies, typically many of which are zero.  On account of these zero components of $\zs$, the probability vector cannot be expressed in the form of Equation \eqref{zrep} and, consequently, $\KL{\zs}{g_d(\vbeta)}>0$, for all $\vbeta\in\R^{t_d}$ and $d\in\{1,2,\ldots,p\}$.  Moreover, let $\ms_d=\T_d^\prime\zs$ be the vector of sample moments up to $d^{th}$-order and note that if there exists a zero component in $\ms_d$, then the system in Equation \eqref{score} does not have a solution $\hat{\vbeta}\in\R^{t_d}$.
These observations indicate that in order to estimate the $d^{th}$-order solution $\hat{\vbeta}\in\R^{t_d}$ with a method modified from that of the preceding section, it is necessary that the first $t_d$ elements of $\zs$ be strictly positive.  More generally, we define the regularized $d^{th}$-order cross-entropy optimization problem:
\begin{align}\label{reg}
\mbox{minimize}\hs H(\zs,g_d(\vbeta))+\lambda\vbeta^\prime\W\vbeta\hs\mbox{for }\vbeta\in\R^{t_d},
\end{align}
for some hyperparameter $\lambda>0$, where $\W=\diag{\w}$ and $\w\in\R^{t_d}$ with element
$w_j =[(d-1)\binom{p}{|u(\x_j)|}]^{-1}$, for $j\in\{p+1,\ldots,t_d\}$ and $w_j=0$ for $j\in\{1,\ldots,p\}$, so that $\|\w\|_1=1$ and the weights $\w$ are such that the regularization $\lambda\vbeta^\prime\W\vbeta$ is order-wise equally applied to the value of the objective function, except to the first-order which is not regularized.

The regularized $d^{th}$-order objective function in Equation \eqref{reg} has corresponding gradient
\begin{align*}
\frac{d}{d\vbeta}\left[H(\zs,g_d(\vbeta))+\lambda\vbeta^\prime\W\vbeta\right] &=
\T_d^\prime[g_d(\vbeta)-\zs]+2\lambda\W\vbeta,
\end{align*}
for some $d\in\{2,3,\ldots,p\}$ and $\lambda>0$.  The solution
\begin{align}\label{regsol}
\hat{\vbeta}_{d\lambda} &= \underset{\vbeta\in\R^{t_d}}{\arg\min}\hs H(\zs,g_d(\vbeta))+\lambda\vbeta^\prime\W\vbeta
\end{align}
to the regularized optimization problem in Equation \eqref{reg} is readily obtained via gradient descent and the regularization parameter $\lambda>0$ is selected via leave-one-out cross-validation (see Appendices \ref{app:grad} and \ref{app:cv}).  We denote $\hat{\z}_{d}=g_d(\hat{\vbeta}_{d\lambda})$, i.e., without the $\lambda$ subscript, as the estimated probability distribution that results from an automatically selected $\lambda$ via the cross-validation subroutine.

\subsection{Information Content of High-order Associations}

We seek to quantify the statistical information attributable to $d^{th}$-order associations in the statistical distribution $\hat{\z}_{d}=g_d(\hat{\vbeta}_{d\lambda})$, for $d\in\{2,3,\ldots,p\}$.  Understanding entropy as a measure of the lack of predictability of states based on a statistical representation of their respective likelihoods,
the Kullback-Leibler divergence
from the estimated $d^{th}$-order probability distribution $\hat{\z}_d$ to the empirical probability distribution of the data $\zs$ is the information discrepancy $\Ical_d = \KL{\zs}{\hat{\z}_d}$.  For values of $d\in\{1,2,\ldots,p\}$, we define the sequence
\begin{align}
\Ical_d &= \label{Ical} -\sum_{j=0}^{\twop-1} \zse_j\left[\log(\hat{z}_{dj})-\log(\zse_j)\right]\\
&= \nonumber H(\zs,\hat{\z}_d)-H(\zs),
\end{align}
where $H(\zs)=-\sum_{k=0}^{\twop-1}\zse_k\log(\zse_k)$ is the entropy of $\zs$.  Note that
\begin{align*}
\Ical_1 &= \left[\Ical_1-\Ical_p\right]+\Ical_p
\end{align*}
decomposes into (i) $\Ical_1-\Ical_p$, 	a global measure of statistical dependence \citep{Rothstein52} known as the mutual information from the first-order $\hat{\z}_1$ distribution to the fully-specified $\hat{\z}_p$ probability distribution, and (ii)
$\Ical_p$ is the lack of fit in approximating $\zs$ with $\hat{\z}_p$ on account of the inexpressibility of $\zs$ in the form of Equation \eqref{zrep}.  The information gained from approximating $\zs$ with the estimated $d^{th}$-order distribution $\hat{\z}_d$ relative to approximating $\zs$ with the $(d-1)^{th}$-order distribution $\hat{\z}_{d-1}$ is
\begin{equation*}
\Ical_{d-1}-\Ical_d = -\sum_{j=0}^{\twop-1}\zse_j\left[\log(\hat{z}_{(d-1)j})-\log(\hat{z}_{dj})\right],
\end{equation*}
for $d\in\{2,3,\ldots,p\}$, and is interpreted as the quantity of statistical information encoded by $d^{th}$-order statistical associations detected in the system with empirical probability distribution $\zs$.  This information quantity is a proportion
\begin{align}\label{reta}
\Rcal_d &= \left(\frac{\Ical_{d-1}-\Ical_d}{\Ical_1-\Ical_p}\right)\in[0,1]
\end{align}
of the total quantity of detected statistical information contained in associations of all orders, for $d\in\{2,3,\ldots,p\}$.  Clearly, $\sum_{d=2}^p \Rcal_d =1$.  Finally, let
\begin{align}\label{rstar}
\Rast &= \frac{\Ical_1-\Ical_p}{\Ical_1}=1-\frac{\Ical_p}{\Ical_1}
\end{align}
be the measure of the quality-of-fit in approximating $\zs$ with the distributions $\hat{\z}_d$ of the form in Equation \eqref{zrep}, for any $d\in\{2,3,\ldots,p\}$. This quantity may be intuited as an information-analogue to the coefficient of determination in linear models.  Its value is attributed to (i) the inexpressibility of the empirical probability distribution in the form of Equation \eqref{zrep} and (ii) the regularization involving $\lambda>0$.  The contributions of each type to the $\Rast$ statistic may be diminished in the large sample limit.  For example, provided that the unknown, true probability distribution of the states in the system is stationary and strictly positive, i.e., is representable in the form of Equation \eqref{zrep}, a sufficiently-large sample size is necessary to include observations on all possible states so that $\Rast=1$ with high probability.  Conversely, the leave-one-out cross-validation procedure have, in the large sample limit, more similarly distributed training data sets and a correspondingly smaller $\lambda$ value on problems of fixed size.

\subsection{Assessing Method Accuracy via Simulation}

\newcommand{\Jcal}{\mathcal{J}}

In the following, we develop a framework for simulating data and assessing the accuracy of our estimation procedure.  Recall that the quantity of primary concern is the information ratio $\Rcal_d$, for $d\in\{1,2,\ldots,p\}$, as defined in Equation \eqref{reta}.  The consistency of this quantity is a function of both (i) our estimation procedure and (ii) the variability in the data set.  In this section, we seek to estimate the consistency of our estimator $\R_d$ of the proportion of information attributed to the $d^{th}$-order statistical associations in the binary system.  To that end, we set $p=5$ and, since third-order associations are the point of departure from a linear to a nonlinear model and are the foremost topic of this document, we set $d=3$ in the present simulation study.

A single pass of our simulation proceeds according to the following.  We set $d=3$ and select a subset of $p=5$ ASV occurrence variables from our data set in the manner presented in the Results section.  Subsequently, we estimate the coefficients $\vbeta_3$ in accordance with the Results section and treat this vector of coefficients as \textit{known} and compute the corresponding probability distribution $\z_3=g_3(\vbeta_3)$.  Based on this probability vector, $B=\t$ independent random samples of $n$ rows, in which $n$ rows are randomly sampled with replacement from the $2^p=32$ rows of the binary matrix $\T_1$, are selected to create random samples $\tilde{\T}_1^{(i)}$, for $i=1,2,\ldots,\t$.  We select a range of values for $n$ to assess the variability of our estimator as a function of sample size.

Corresponding to this collection of $B=\t$ random samples and for a particular sample size $n$, we estimate $\hat{\vbeta}_3^{(i)}$
in accordance with the Methods section and subsequently estimate the information quantity $\Rcal_3^{(i)}$, for each random samples $i=1,2,\ldots,\t$.

Using our previously outlined methods, we estimate $\Rcal_3^{(i)}$ and compare the resulting estimates to the ground truth $\Rcal_3$, as computed from the original sample, and record the median and interquartile range of the deviations from $\Rcal_d^{(i)}$ to $\Rcal_d$, for $i=1,2,\ldots,\t$, in Table \ref{tab} (see Appendix \ref{app:sim} for more information and histograms of the simulated results).  Over these sets of simulations, we set the sample size $n=2^k\cdot381$, for $k=0,1,2,3$, in order to demonstrate the statistical performance of our estimator $\Rcal_d$ as a function of sample size.

\begin{table}[!ht]
	\begin{tabular}{||r||c|c|c||}
		\hline\hline
		$n$ & Median  & IQR & IQR width\\
		\hline\hline
		$381$ & -0.005  & (-0.041, 0.100) & 0.141\\
		\hline
		$2\times381$ & -0.001 & (-0.033, 0.096) & 0.129\\
		$4\times381$ & 0.000 & (-0.025, 0.081) & 0.106\\
		$8\times381$ & 0.000 & (-0.019, 0.065) & 0.084\\
		\hline\hline
	\end{tabular}
	\caption{\textbf{Results of simulation on data replicating original data consisting of five ASV occurrence variables.}  In each instance, we initially randomly select a subset of $p=5$ ASV binary occurrence variables from the original data set and estimate $\vbeta_3$ on this data set.  We use this $\vbeta_3$ vector to compute the estimated probability vector $\z_3$ and randomly sample with replacement $n$ observations from the rows of $\T_1$ and subsequently estimate $\Rcal_3$.  We treat this value as the ground truth.  This process is repeated $B=\t$ times on subsets of $p=5$ ASV binary occurrence variables for a total of $\t$ estimation instances.  We compute $\Rcal_3$ on each of these $20\times B$ subsets of $p=5$ subsets and record the median of the residuals between the estimated $\R_3$ quantities and the ground truth values along with the interquartile range (IQR) of these residuals over all simulated samples of size $n=2^k\cdot381$, for $k=0,1,2,3$.  The corresponding histograms depicting the sampling distribution of the residuals of our estimator are presented in Appendix \ref{app:sim}.}
	\label{tab}
\end{table}
The statistics in Table \ref{tab} that describe the results of our simulation study are evidence that, on the simulated data that mimics our original data, our estimator is accurate.  These results are further supported by the supplemental information that we provide in Appendix \ref{app:sim} along with analogous results for the cases that $p=8$ and $p=10$.

\section{Results}

To provide further context for our forthcoming results on the complexity of associations of the gut microbiota, we begin by providing insight on stereotypical nonlinear statistical associations and the performance of our methodology in these situations and subsequently we present the results of our analysis on the infant gut microbiome data collected at the New Hampshire Birth Cohort.

\subsection{Example}

A nonlinear association among three binary variables $x_1,x_2,x_3$ is exemplified through the standard \textit{exclusive or}, i.e., ``at least one but not both'', example \citep{Barbour1992} in which the probabilities $p_{00}=\Pr{x_1=0,x_2=0}$ and $\Pr{x_1=x_1^\prime,x_2=x_2^\prime}=(1-p_{00})/3$, for $(x_1^\prime,x_2^\prime)\neq (0,0)$ and $x_3=x_1+x_2-2x_1x_2$, so that $x_3=1$ when either $x_1=1$ or $x_2=1$ and, otherwise, $x_3=0$.  Note that the three variables are equiprobable with $\Pr{x_j=1}=2(1-p_{00})/3$ and equicorrelated with $\mbox{Cov}(x_i,x_j)=(4p_{00}-1)(1-p_{00})/9$,	
for $i,j\in\{1,2,3\}$ and $i\neq j$.  If $p_{00}=2/5$, then the three variables are positively correlated with $\mbox{Cor}(x_i,x_j)=1/6$ and, yet, $\Pr{x_1=1,x_2=1,x_3=1}=0$.
The mathematical relationship between $x_3$ and $x_1,x_2$ conceptualizes a nonlinear ecological association in which a triad of distinct ecological units do not occur collectively but are, otherwise, supportive of pairwise co-occurrences (see
the Supporting Information).

\begin{figure}[!ht]
	\includegraphics[width=.9\linewidth]{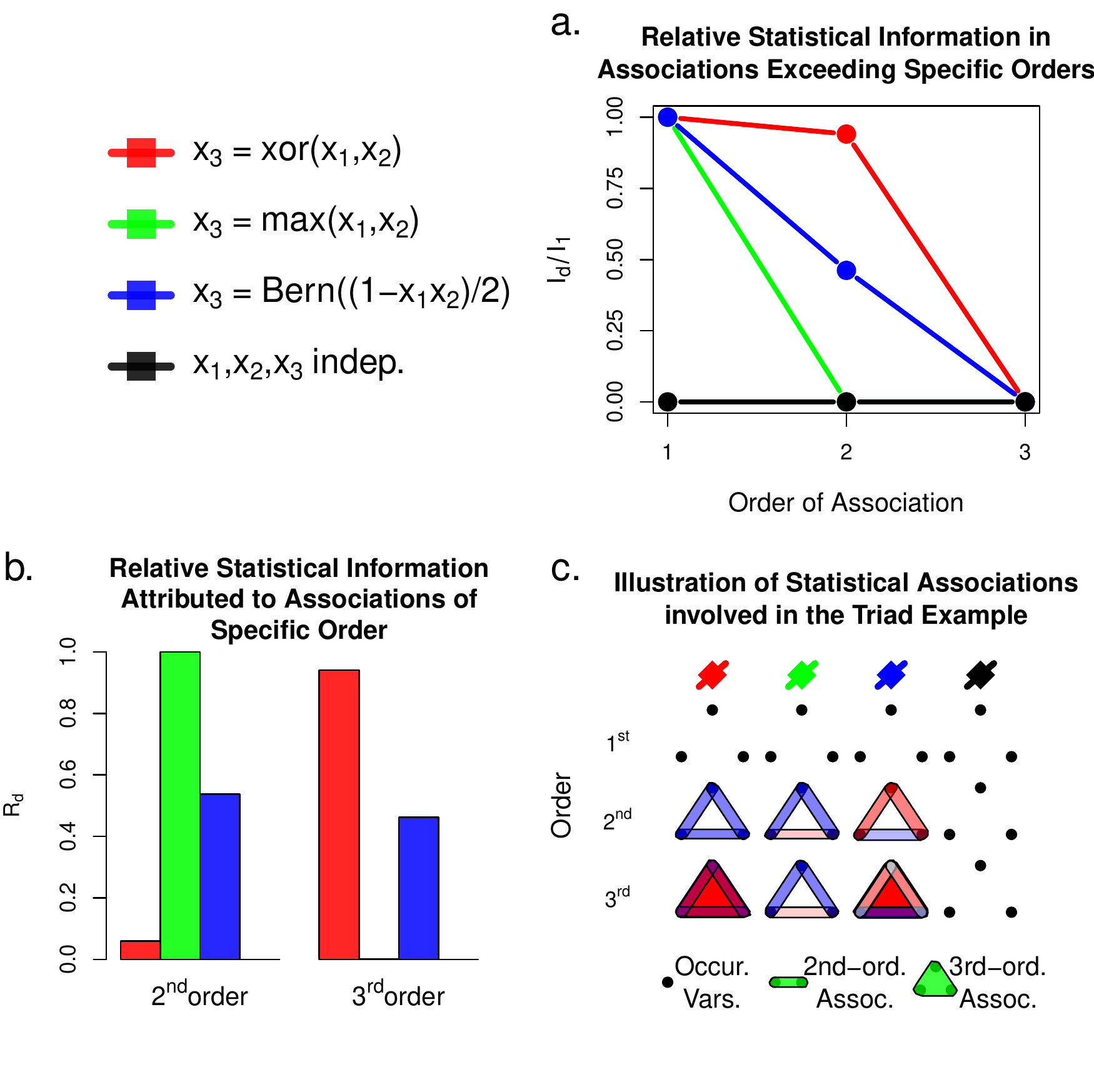}
	\caption{\textbf{Illustration of the information content of third-order associations.}
		For three occurrence variables $x_1,x_2,x_3$, with $\Pr{x_1=0,x_2=0}=2/5$ and $\Pr{x_1=x_1^\prime,x_2=x_2^\prime}=1/5$, for $(x_1^\prime,x_2^\prime)\neq (0,0)$, we define the models (red) ``exclusive or'' (\textit{xor}) $x_3=x_1+x_2-2x_1x_2$, (green) \textit{max} $x_3=\max\{x_1,x_2\}$, (blue) \textit{relaxed xor} $x_3=0$, if $x_1=1$ and $x_2=1$, and, otherwise, $x_3\sim\mbox{Bernoulli}(1/2)$, denoted as $\textit{Bern}$ in the figure legend, and (black) $x_1,x_2,x_3$ mutually independent \textbf{a.} For each $d=1,2,3$, node points on the interpolated plotted points represent the statistical information attributable to orders $>d^{th}$-order as a proportion $\Ical_d/\Ical_1$ (see Methods) of information attributable to all orders of association.  \textbf{b.} Proportion of statistical information attributable to $d^{th}$-order associations $\Rcal_d$, for $d=2,3$.  \textbf{c.} Visualization of associations represented in the maximum entropy distribution approximation to the distribution of states in the subsystem of three occurrence variables, denoted as \textit{Occur. Vars.} in the figure, from each model.  A relatively strong association is shaded darker than weaker associations.  Positive associations are shaded blue, whereas negative associations are shaded red.}
	\label{fig:one}
\end{figure}

A linear classifier, e.g., logistic regression, for the occurrence variable $x_3$ based on the main variables $x_1$ and $x_2$ (no interaction term) would be expected to incorrectly classify $x_3$ with rate $\Pr{x_1=1,x_2=1}=(1-p_{00})/3$ which, in the present illustration is $\Pr{x_1=1,x_2=1}=1/5$, for example, when trained on a random sample of subsystem states (see the Supporting Information). 
To that end, the importance of the inclusion of third-order association among the three occurrence variables $x_1,x_2,x_3$ in a statistical representation of the occurrence states is depicted in Fig.  \ref{fig:one}a.  For the sake of illustration, we include the \textit{max} model in which $x_3=\max\{x_1,x_2\}$ and the \textit{relaxed xor} model in which $x_3$, conditional on $x_1,x_2$, has the $\mbox{Bernoulli}\left(\frac{1-x_1x_2}{2}\right)$ distribution.

Each interpolation of the plotted points $(d,\Ical_d/\Ical_1)$, where $\Ical_d$, for $d\in\{1,2,3\}$, graphs the proportion of statistical information attributed to associations higher than $d^{th}$-order as measured by the Kullback-Leibler divergence \citep{Kullback1951}, a measure of relative entropy (see Methods) to the true probability distribution $\z$ of the three variables from the $d^{th}$-order approximating maximum entropy distribution $\z_d$.  For all data sets in general, the paths described are non-increasing functions of order $d\in\{1,2,\ldots,p\}$.  The difference $\Rcal_d\propto \Ical_{d-1}-\Ical_d$ is reflected in the negative magnitude of the jumps.

With respect to the \textit{xor} model, the second-order maximum entropy distribution in the form of Equation \eqref{zrep} is a modest improvement, accounting for an $\approx 6\%$ reduction, in the relative information from its first-order counterpart to the true probability distribution of the \textit{xor} model.  This carries the interpretation that the second-order probability model is expected to be approximately equally predictive of occurrence states as the first-order probability model.  However, with the third-order association encoded, the third-order probability model approximates the true probability distribution of states with arbitrary accuracy.  The second- and third-order associations present in the third-order model encode all of the statistical information attributable to the associations of this subsystem.  While this is trivially the case since we encoded associations of all possible orders in this third-order probability model, it is straight-forward that the argument would remain the same, for example, in the context of a fourth and fully-independent occurrence variable $x_4$.  Accordingly, the maximal order of four would need not be encoded to represent all statistical information attributable to associations in the expanded subsystem.

The fully independent model encodes no associations and is provided as a baseline for comparison.  The \textit{max} model is constructed, like the \textit{xor} model, to define $x_3$ deterministically as a function of $x_1,x_2$ but to alleviate the necessity to encode a third-order association to represent the full statistical information of all orders.  The \textit{relaxed or} assigns zero probability to the event $x_j=1$, for $j=1,2,3$, and, otherwise, is a $\mbox{Bernoulli}(1/2)$ random variable negatively associated pairwise with $x_1,x_2$.  A visualization of the existence, sign, and magnitude of the associations of the three different models in each of the approximating maximum entropy distributions of orders $d\in\{1,2,3\}$ is seen in Fig.  \ref{fig:one}c.  Selected mathematical details are provided in the Supporting Information. 

\subsection{Analysis of Infant Microbiota Data}

To estimate the statistical information quantities of interest in subsystems of $p$ amplicon sequence variants (ASVs) ecological occurrence variables, for $p\in\{5,8,10\}$, we obtain a random sample of $\t$ subsets of $p$ ASVs from the set of all $\binom{\pp}{p}$ such subsets and record their respective ASV occurrence profiles, i.e., observed ecological states, over all $\n$ infant samples (see Methods).  For each subset of $p$ ASVs, we estimate the $d^{th}$-order maximum entropy model $\hat{\z}_d=g_d(\vbeta_{d\lambda})$ (see Methods) and compute
\begin{align*}
\Rcal_d &= \frac{\Ical_{d-1}-\Ical_d}{\Ical_1-\Ical_p},
\end{align*}
for $d\in\{1,2,...,5\}$.\footnotemark\footnotetext{We compute $\Rcal_\ast=1-\Ical_p/\Ical_1$ for $d=5$ (see Appendix \ref{app:resid}).}  We consider at most $5^{th}$-order representations due to the sample size and computational limitations related to the numerical estimation of $\twop-1=\sum_{j=1}^p\binom{p}{j}$
parameters in the optimization problem in Equation \eqref{reg}. 

\begin{figure}[!ht]
	\includegraphics[width=.9\linewidth]{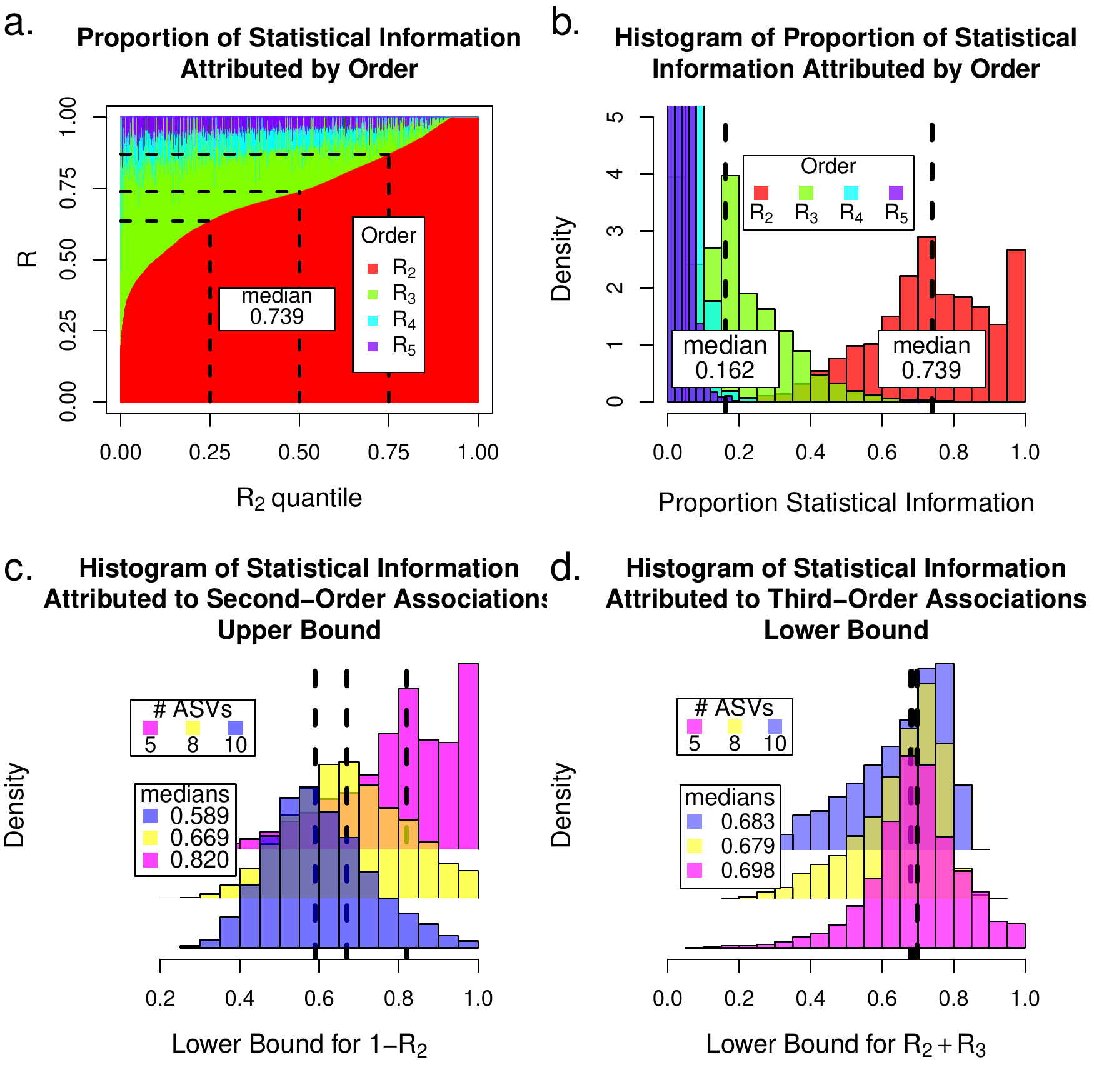}
	\caption{\textbf{Results of nonlinear association inference in subsystems of microbiota ecological occurrence variables as identified through ASVs.}
		\textbf{a.} Over a random sample of \t subsets of $p=5$ ASVs, the proportion of statistical information attributed to each order $\Rcal_d$, for $d\in\{2,3,4,5\}$, arranged according to the sorted values of $\Rcal_2$ with quartiles indicated.  The sample median $q_{0.5}=\twoprop$ of the $\Rcal_2$ statistic indicates that the estimated second-order associations account for at most 75\% of statistical information attributable to all orders of association in no more than half of the sampled five-vertex subsets.  \textbf{b.} Over the same set of \t subsets of $p=5$ ASVs, histograms of the proportion of statistical information attributed to each order $\Rcal_d$, for $d\in\{2,3,4,5\}$. \textbf{c.} Upper bound for $\Rcal_2$, the proportion of statistical information attributed to second-order associations, for subsets of $p\in\{5,8,10\}$ ASVs, over a random sample of \t subsets each.  \textbf{d.}  Lower bound for $\Rcal_2+\Rcal_3$, the proportion of statistical information attributed to third-order associations, respectively, over the samples.}
	\label{fig:two}
\end{figure}

On subsystems of $p=5$ ASV occurrence variables, we estimate a median \twoprop proportion of statistical information content attributed to second-order associations $\Rcal_2$ and an interquartile range of \IQR over the $\t$ subsets.  Fig.  \ref{fig:two}a illustrates this prominent role for second-order associations in the prediction of occurrence states across subsets.  The $\eqtwo$ of subsets, which have statistical information allocated entirely in the second-order associations, induce the bimodal distribution illustrated in Fig.  \ref{fig:two}b.  The remaining statistical information is primarily attributable to third-order associations, the median \threeprop proportion of statistical information attributed to third-order associations $\Rcal_3$.  The median value \compthree of $\Rcal_2+\Rcal_3$ over all \t subsets of $p=5$ ASV occurrence variables and an interquartile range of \IQRthree indicates that the vast majority of statistical information pertinent to prediction of ecological states is attributable to second- and third-order associations encoded in the third-order maximum entropy models.  The $\Rast=1-\Ical_p/\Ical_1$ quality-of-fit statistic (see Methods) has a median value \rastmed with a corresponding interquartile range \rastiqr. This implies that our model estimation procedure encoded (through associations of all orders) at least \rastlo of the mutual information among $p=5$ ASVs occurrence profiles in 75\% of \t samples.  The complementary quantity, i.e., the other part of the whole, is the residual information that is attributable to a finite sample size and regularization (see Methods and Appendix \ref{app:resid}).

The third-order maximum entropy models for $p=8$ and $p=10$ occurrence variables are parameterized by $\sum_{j=1}^3\binom{8}{j}=92$ and $\sum_{j=1}^3\binom{10}{j}=175$ quantities, respectively, whereas by comparison, $31$ parameters are involved in the fully-specified model for $p=5$ variables.  In lieu of computing $\Ical_p$, a quantity specified by $2^8-1=255$ and $2^{10}-1=1023$ parameters for $p=8$ and $p=10$, respectively, we bound
\begin{align}
\frac{\Ical_1-\Ical_2}{\Ical_1} \leq \Rcal_2 &= \label{ineqone} \frac{\Ical_1-\Ical_2}{\Ical_1-\Ical_p} \leq \frac{\Ical_1-\Ical_2}{\Ical_1-\Ical_3}\\
\frac{\Ical_2-\Ical_3}{\Ical_1} \leq \Rcal_3 &= \label{ineqtwo} \frac{\Ical_2-\Ical_3}{\Ical_1-\Ical_p} \leq \frac{\Ical_2-\Ical_3}{\Ical_1-\Ical_3},
\end{align}
since $0\leq\Ical_p\leq\Ical_3$, for $p=8,10$.  Adding the two inequality systems in Equations \eqref{ineqone} and \eqref{ineqtwo} results in
\begin{align*}
1-\frac{\Ical_3}{\Ical_1} &\leq \Rcal_2+\Rcal_3 \leq 1.
\end{align*}
In Fig.  \ref{fig:two}c, we note that the median upper bound for $\Rcal_2$ for each of the $p\in\{5,8,10\}$ considered is evidently decreasing with $p$ and indicates a diminished prediction efficacy of the second-order maximum entropy models and necessarily greater quantities of statistical information attributed to higher associations in the probability models on increasingly larger sub-ecosystems.  The median lower bound $1-\Ical_3/\Ical_1$ for the combined statistical information encoded in the second- and third-order maximum entropy models are \medfive, \medeight, and \medten, for $p=5,8,10$, respectively.  Fig.  \ref{fig:two}d illustrates empirically that the lower bound statistic for $p=5,8,10$ are distributed similarly, each with an interquartile range of approximately $\IQRapx$.  This common statistical behavior of the estimated lower bound for the quantity of statistical information pertinent to prediction of occurrence states in small sub-ecosystems attributable to the associations encoded in the second- and/or third-order maximum entropy models predominates the total quantity of statistical information attributed to all orders of association.  In other words, second- and third-order associations are the foundation of an accurate representation of the collective statistical behavior of microbial ecological occurrence variables in small sub-ecosystems of the infant gut. 

\section{Discussion}\label{discuss}

The states of a microbial sub-ecosystem of the infant gut elicit a statistical description which encodes high-order associations.  In our manner of estimating and allocating the statistical information attributed to the associations encoded in a sequence of maximum entropy models of increasing specification in our observations of the infant gut microbiota ecosystem, we identified an \strike{prominent} \blue{influential} role for the third-order maximum entropy model, as a statistical representation of sub-ecosystems, in the prediction of ecological states.  Third-order associations predominate the estimated high-order associations. In concert with the pairwise association they may encompass, second- and third-order associations are typically ascribed at least two-thirds of the information attributable to associations of all orders of ecological occurrence variables.  Our statistical analysis provides convincing evidence that small subsystems of $5\leq p\leq 10$ infant microbiota ASV ecological occurrence variables are commonly at least third-order systems.

Our methodology for attributing the components of statistical information to specific orders of association is widely applicable within complex systems analysis.  In particular, our estimation and evaluation procedure gathers evidence from the data to quantify the predictability of states as a function of model complexity which, in the present context, is indexed by the maximal order of association in a maximum entropy distribution approximating the empirical distribution of the observed states.

Our subsequent analysis \strike{uncovered the existence of and} quantified the extent to which associations among microbial occurrence variables are nonlinear.  The accurate prediction of outcomes from clinical interventions or perturbations of the gut ecosystem \blue{may warrant} \strike{will require} encoding these complex relationships of the system.  We have established the breadth of high-order associations that modify the lower-order association they encompass in the gut microbiota environment and, as a result, founded a baseline for the level of difficultly in prediction of gut microbiota states.

\blue{We have elected to focus on the most inherently unpredictable ASVs to estimate a lower bound for the predictability of the entire collection of ASVs that were identified during the sequencing process.  While our results may only identify some community differences when compared to traditional methods, we have primarily identified differences in predictability and stability within communities.  For example, a triad of ASVs, each pairwise positively-associated with the others, represents a subsystem of runaway positive feedback.  Conversely, with a negative triplewise association overlaying this triad, the subsystem is governed and potentially inferred to be stable.  Ultimately, these higher-order associations may be viewed as moderating forces and are potentially the structure required to stabilize the community in the event of a disturbance to the gut microbiota system.}


We have estimated and quantified the essential statistical behavior of ecological co-occurrences of microorganisms in the infant gut microbiota  in terms of their mutual associations.
Based on $\n$ samples collected from participating infants, we demonstrated that \blue{ecological states of} small (up to 10 ASVs) subsystems \blue{of} the infant gut ecosystem \strike{predominantly require} \blue{commonly exhibit high-order associations. These associations warrant} a probability model which represents the statistical properties of third-order, i.e., involving three units, ecological co-occurrences since\blue{, when compared to basic second-order graphical models, the more expressive models frequently encoded} the statistical information attributable to high-order ecological associations that are relevant to the accurate prediction of ecological states.  We described our numerical procedure for the estimation of the proportion of statistical information ascribed to a range of orders of association in a binary representation of the co-occurrences of units in an ecosystem.

\section{Conclusion}

The statistical behavior of ecological co-occurrences among the microorganisms of the infant gut microbiota is vital information for accurate prediction of possible states of the ecosystem.
By applying our methodology to data collected from infants participating in the New Hampshire Birth Cohort Study,
we demonstrated that \blue{statistical associations among ecological variables in} small subsystems \blue{of} the infant gut microbial ecosystem \strike{predominantly require} \blue{are regularly nonlinear.  Accordingly,} a probability model which \strike{encodes} \blue{specifies} the statistical properties of third- \blue{and higher-order association among} co-occurrences of ecological units, \blue{is necessary} \strike{in order} to \blue{completely} represent the statistical information attributable to ecological associations and relevant to prediction of ecological states.  \blue{We hypothesize that, in general, the order of statistical association of ecological subsystems is a function of the tree depth of the corresponding tropic network.  Nevertheless, we leave for future work this characterization of the gut microbiota among other human body sites.}  In summary, we described our numerical procedure for the estimation of the proportion of statistical information ascribed to a range of orders of co-occurrences of ecological units in an ecosystem and propose that our findings imply an important role for complex interactions of microbes in the human infant gut.

\begin{appendix}
	

\section{Gradient Descent}\label{app:grad}

The gradient descent update
\begin{align}\label{grad}
\vbeta^{(1)} &= \vbeta^{(0)}-\T_d^t\left[g_d(\vbeta^{(0)})-\z\right]
\end{align}
is our preferred method for numerical optimization in this context. Strict positivity of $\z$ is sufficient for the above iteration to converge to the optimum $\hat{\vbeta}_d\in\R^{t_d}$ such that $H(\z,\hat{\z}_d)<H(\z,\z_d)$, for any $\z_d\in\zotp$, where $\hat{\z}_d=g_d(\hat{\vbeta}_d)$.  For general $\vbeta\in\R^{t_d}$ and $\z_p=g_d(\vbeta)$, the gradient is expressed as
\begin{align*}
\frac{d}{d\vbeta}g_d(\vbeta) &= \frac{d}{d\vbeta}\expl{\T_d\vbeta-\log(Z(\vbeta))\one}\\
&= \left[\T_d^\prime-
\T_d^\prime\cdot\z_p\cdot\one^\prime\right]\mbox{diag}(\z_p)\\
&= \T_d^\prime\left[\I_{\twop}-\z_p\cdot\one^\prime\right]\diag{\z_p},
\end{align*}
which implies that
\begin{align*}
\frac{d^2}{d\vbeta d\vbeta^\prime}H(\z,g_d(\vbeta)) &= \frac{d}{d\vbeta}\left[\T_d^\prime g_d(\vbeta)-\z\right]\\
&= \T_d^\prime\left[\I_{\twop}-\z_p\cdot\one^\prime\right]\diag{\z_d}\T_d\\
&= \T_d^\prime\left[\diag{\z_d}-\diag{\z_p}\J_{\twop}\diag{\z_p}\right]\T_d\\
&= \T_d^\prime\Rbf\T_d,
\end{align*}
where $\R=\diag{\z_d}-\diag{\z_p}\J_{\twop}\diag{\z_p}$.  Note that the $k^{th}$ diagonal element $R_{kk}=z_k(1-z_k)$ and the sum of the off-diagonal elements of the $k^{th}$ row of $\R$ is
\begin{align*}
\Rbf_{k\cdot} &= -z_{dk}\sum_{\substack{j=0\\j\neq k}}^{\twop-1}z_{dj}\\
&= z_{dk}^2-z_{dk}\sum_{j=0}^{\twop-1}z_{dj}\\
&= z_{dk}(1-z_{dk}),
\end{align*}
which demonstrates that $\Rbf$ is weakly diagonally dominant with non-negative diagonal elements and, thus, is positive semi-definite.  In particular, since $\Rbf\one=\zero$, then $\v=\one$ is the eigenvector of $\R$ with eigenvalue $\lambda=0$.  Since no column of $\T_d$ is proportional to $\one$ and the columns of $\T_d$ are linearly independent, then it follows that $\T_d^\prime\Rbf\T_d$ is positive definite and $H(\z,g_d(\vbeta))$ is a convex function of $\vbeta$.

\section{Cross-Validation}\label{app:cv}

\renewcommand{\S}{\mathbf{S}}

Our cross-validation procedure assesses an information-based measure of discrepancy from estimated leave-one-out probability distributions to the empirical probability distribution over a sequence of regularization parameter values.

Viewing $\zs$ as a probability vector over the integers $\S_p=\{0,1,\ldots,\twop-1\}$, define $\K=\{k\in\S_p:\zse_k>0\}$ as the integers corresponding to the binary representations of the observed state vectors obtained in the sample.  Let $\zs^{(k)}\in\zotp$ be the empirical probability distribution vector obtained after removing one observation corresponding to some $k\in\K$.  The $j^{th}$ element of $\zs^{(k)}$ is
\begin{align*}
\zse_j^{(k)} &= \frac{n}{n-1}\left(\zse_j-\frac{1}{n}1\{j=k\}\right),
\end{align*}
for $j\in\{0,1,\ldots,\twop-1\}$.  The probability vector $\hat{\z}_{d\lambda}^{(k)}=g_d(\hat{\vbeta}_{d\lambda}^{(k)})$ estimate based on $\zs^{(k)}$ is identified by
\begin{align}
\hat{\vbeta}_{d\lambda}^{(k)} &= \underset{\vbeta\in\R^{t_d}}{\arg\min}\hs H(\zs^{(k)},g_d(\vbeta))+\frac{\lambda}{2}\|\w\circ\vbeta\|_2^2.
\end{align}
Let $\hat{z}_{d\lambda k}^{(k)}$ be the $k^{th}$ element of $\hat{\z}_{d\lambda}^{(k)}$ and define
\begin{align}\label{cv}
\hat{\lambda}_d &= \underset{\lambda>0}{\arg\min}\hs -\sum_{k\in\K}\frac{\zse_k}{\sum_{j\in\K}\zse_j}\log\left(\frac{\hat{z}_{d\lambda k}^{(k)}}{\sum_{j\in \K}\hat{z}_{d\lambda j}^{(j)}}\right)
\end{align}
as the $\lambda>0$ value that minimizes the loss function in Equation \eqref{cv} which is a generalization of cross-entropy between the empirical probability vector $\zs$ and the collection of leave-one-out probability vector estimates $\hat{\z}_{d\lambda}^{(k)}$, for $k\in\K$.  That is, the optimal regularization parameter $\hat{\lambda}_d$ is selected as the non-negative value which minimizes, up to a constant, the average information loss in approximating the $k^{th}$ element $\zse_k$ of the empirical distribution of the data $\tilde{\z}$ with the $k^{th}$ component of the leave-one-out estimated distributions $\hat{z}_{d\lambda k}^{(k)}$, over all $k\in\K$.  The normalization factors in Equation \eqref{cv} are included to standardize the loss function.  Up to the requisite normalization factors, the probability $\zse_k$ is paired with $\hat{z}_{d\lambda k}^{(k)}$, i.e., the sample value against its corresponding leave-one-out value.  The regularized parameter estimate $\hat{\vbeta}_{d\hat{\lambda}}$ identifies the regularized $d^{th}$-order maximum entropy distribution $\hat{\z}_{d\hat{\lambda}_d}=g_d(\hat{\vbeta}_{d\hat{\lambda}})$.  Selecting $\hat{\lambda}_d$ is automatic and efficient via a golden ratio search \citep{Kiefer1953}. Accordingly, we suppress the $\lambda$ notation in the $d^{th}$-order maximum entropy distribution $\hat{\z}_d=\hat{\z}_{d\hat{\lambda}_d}$.

\section{Simulation Details}\label{app:sim}

\newcommand{\vgam}{\boldsymbol\gamma}
\newcommand{\vt}{\mathbf{t}}

A component of our simulation procedure is the projection of a $d^{th}$-order maximum entropy distribution $\z_d=g_d(\vbeta_d)$ down to the $(d-1)^{th}$-order maximum entropy distribution. This projection is necessary so that, for example, when we generate a $d^{th}$-order probability vector $\z_d=g(\vbeta_d)$, we are able to determine the corresponding $(d-1)^{th}$-order probability vector and subsequently compute the ground truth $\Rcal_d$ value. This is achieved via the following program.  We have from Equation \eqref{zrep} that
\begin{align*}
\z &= \exp\{\T\vgam-\log(Z(\vgam))\one\}.
\end{align*}
Let's suppose that, in particular,
\begin{itemize}
	\item $\T_d$ is the design matrix for the $d^{th}$-order model and that $\vgam\in\mathbb{R}^{p_d}$ and
	\item $\T_{d^\prime}$ is the design matrix for the $d^{\prime th}$-order model and that $\vbeta\in\mathbb{R}^{p_{d^\prime}}$,
\end{itemize}
for $d^\prime<d$.  This means that
\begin{itemize}
	\item $\z_d=g_d(\vgam)$ is the $d^{th}$-order probability vector and
	\item $\z_{d^\prime}=g_{d^\prime}(\vbeta)$ is the $d^{\prime th}$-order probability vector.
\end{itemize}
Then the cross-entropy
\begin{align*}
H(\z_d,\z_{d^\prime}) &= \sum_{k=0}^{2^p-1} \exp\{\T_d\cdot\vgam-\log(Z(\vgam))\}\left[\vt_{d^\prime k}\cdot\vbeta-\log(Y(\vbeta))\right],
\end{align*}
where $\vt_{d^\prime, k}$ is the $k^{th}$ row of the $\T_{d^\prime}$ matrix and $Z(\vgam)$ and $Y(\vbeta)$ are the partition functions for the $d^{th}$- and $d^{\prime th}$-order maximum entropy distributions, respectively.  Note that the gradient of $H(\z_d,\z_{d^\prime})$ with respect to the parameter vector $\vbeta$ is
\begin{align*}
\frac{d}{d\vbeta}H(\z_d,\z_{d^\prime}) &= -\sum_{k=0}^{2^p-1}\exp\{\vt_{dk}\cdot\vgam-\log(Z(\vgam))\}\left[\vt_{d^\prime k}-\left(\sum_{j=0}^{2^p-1}\frac{1}{Y(\vbeta)}\exp\{\vt_{d^\prime j}\cdot\vbeta\}\vt_{d^\prime j}\right)\right]\\
&= \T_d^\prime\left[\z_d-\z_{d^\prime}\right].\\
\end{align*}
It follows that the solution $\hat{\vbeta}$ to the system of equations $\zero=\T_d^\prime\left[\z_d-\z_{d^\prime}\right]$
may, again, be determined via gradient descent.

The histograms that resulted from our four simulation studies that were initially described in Table \ref{tab}, over $n=2^k\cdot 381$, for $k=0,1,2,3$, are described in the Methods sections and are provided below in Figure \ref{fig:atwo}.  Additionally, we provide statistics from our simulations on subsystems of $p=8$ and $p=10$ variables in Table \ref{tabtwo}.
\begin{figure}[H]
	\includegraphics[width=.9\linewidth]{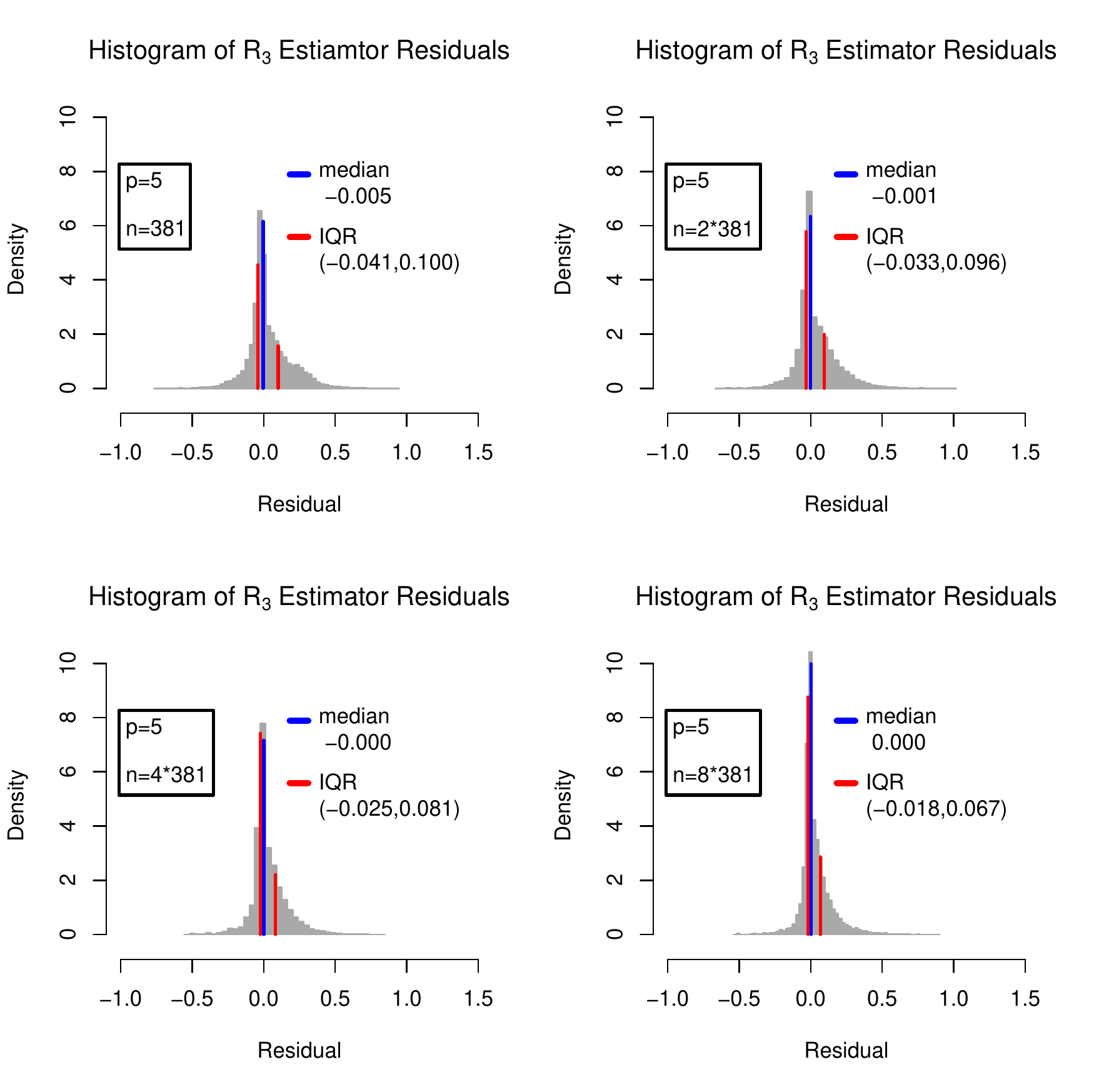}
	\caption{Simulation results of $\Rcal_3$ minus the true proportion of statistical association attributed to third-order association.  That is, the plots above present the residual (difference) between $\Rcal_3$ and the ground truth for the $p=5$ and $n=2^k*381$ context, for $k=0,1,2,3$.  In each instance, the median is computed and plotted with the blue solid line while the lower and upper quartile is plotted with the red line.  The center (median) of the sampling distribution of our estimator of $\Rcal_3$ is a modest underestimate of the actual $\Rcal_3$ value (-0.5\%) but improves to a zero median difference with increasing sample size.  Moreover, the variability, as measured through IQR width, decreases with sample size.}
	\label{fig:atwo}
\end{figure}

\begin{table}[!ht]
	\begin{tabular}{||r||c|c||c||}
		\hline\hline
		& \multicolumn{3}{||c||}{Median}\\
		\hline
		p & Lower Bound & Upper Bound & Width\\
		\hline\hline
		8 & -0.004 & 0.122 & 0.123\\
		10 & -0.009 & 0.201 & 0.191\\
		\hline\hline
	\end{tabular}
	\caption{\textbf{Results of simulation on data replicating original data consisting of eight and ten ASV occurrence variables, respectively.}  We separately and randomly select $p=8$ and $p=10$ subsets of ASV binary occurrence variables from the original data set and estimate $\vbeta_3$ on these data sets.  We use these two $\vbeta_3$ vectors to compute the estimated probability vector $\z_3$ for the $p=8$ and $p=10$ data sets.  In the manner described in the Methods section, we generate $n=\n$ simulated observations and compute the lower and upper bounds in Equations \eqref{ineqone} and \eqref{ineqtwo}.  This process is repeated $B=1000$ times.  We compute the actual $\Rcal_3$ on each of these simulated data sets and record all estimated lower and upper bounds.  The differences between the lower bound and $\Rcal_3$ (lower residual) and the upper bound and $\Rcal_3$ (upper residual) are recorded and presented in the table.}
	\label{tabtwo}
\end{table}

\section{Residual Information}\label{app:resid}

\begin{figure}[H]
	\includegraphics[width=.4\linewidth]{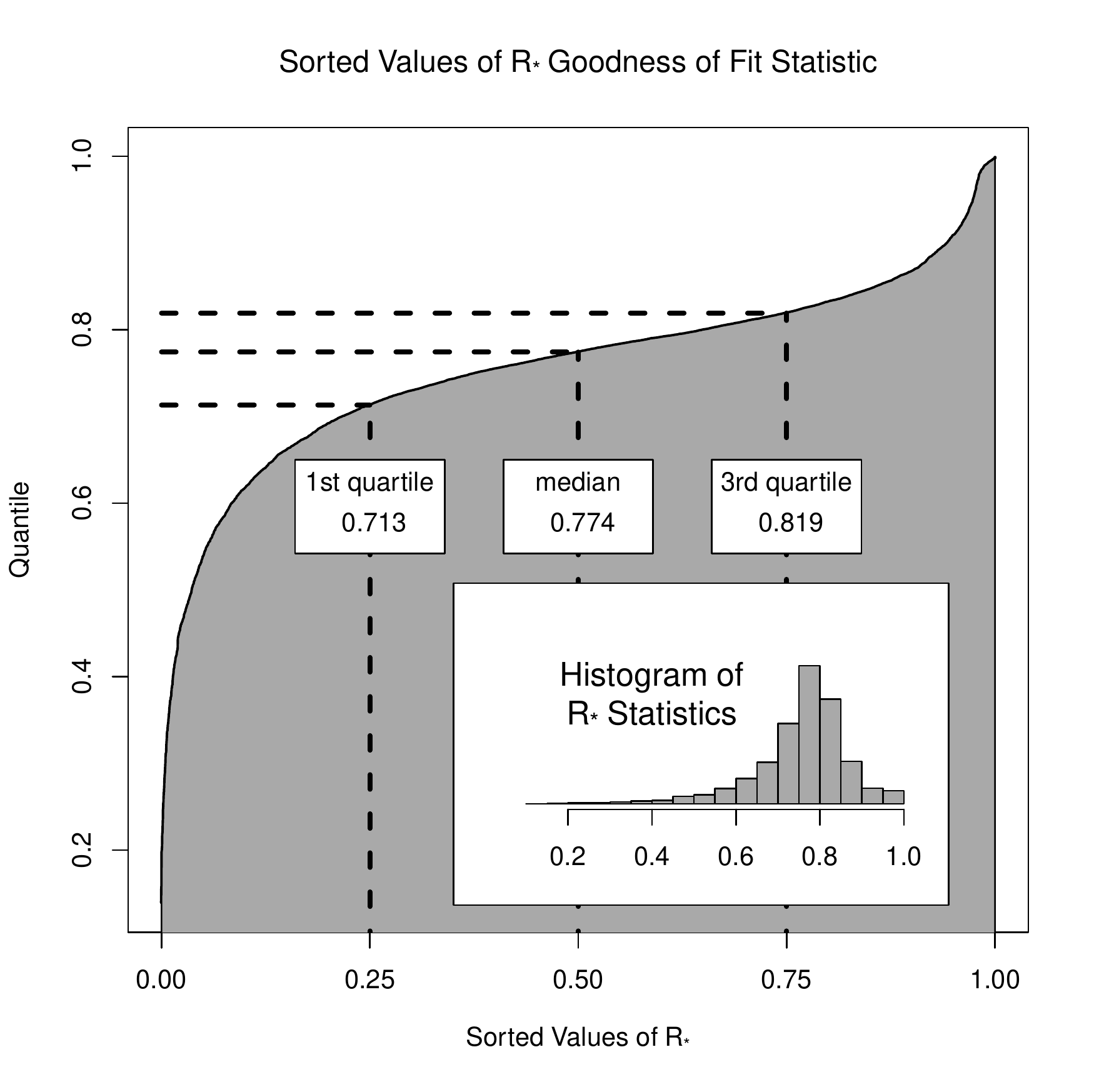}
	\caption{Supplemental. Quantiles of the $\Rast$ statistic on the quality of model fit to the data.}
	\label{fig:si}
\end{figure}

A plot in the format of Figure \ref{fig:two}a, we provide the quantiles of the $\Rast$ quality-of-fit statistic in Figure \ref{fig:si}.

\end{appendix}
%
%

\section*{Acknowledgments}

The authors are grateful to the children and families that made this study possible and to the staff of the New Hampshire Birth Cohort Study.

This work was supported by grants from the National Institutes of Health (OD UG3OD023275, NIEHS P01ES022832, NIEHS P20ES018175, NIGMS R01GM123014, NIGMS P20GM104416, NLM K01LM011985 and NLM R01LM012723) and the US Environmental Protection Agency (RD-83544201 and RD-83459901).

WDV's current affiliation is with the Department of Mathematics and Statistics, University of Southern Maine.

\bibliographystyle{imsart-nameyear} 
\bibliography{AoASbib}       


\end{document}